\documentclass[12pt,epsfig,aps]{article}
\usepackage{graphicx}
\usepackage{epsfig}
\usepackage{amssymb,latexsym,amsmath}
\usepackage[cp1251]{inputenc}

\newcommand{\bs}{\boldsymbol}

\begin{document}

\title{ \Large{ \bf What Can We Learn from Reaction Zone in
Relativistic Nucleus-Nucleus Collisions? } }
\author{ D. Anchishkin$^a$, A. Muskeyev$^b$, S. Yezhov$^b$
\\ $^a$ {\small \it Bogolyubov Institute for Theoretical Physics,
             03680 Kiev, Ukraine}
\\ $^b$ {\small \it Taras Shevchenko Kiev National University,
03022 Kiev, Ukraine}  }

\maketitle

\begin{abstract}
We apply the ``zone of reactions'' as a tool in studying the
interacting system formed in a collision of relativistic nuclei.
With the use of the intensity of collisions of particles (the number
of collisions in unit volume per unit time), we study the space-time
structure of a fireball. In this approach, three basic regions for
the evolution of a system are separated by the scale of the
intensity of collisions: the zone of a hot fireball, the zone of a
cold fireball, and the zone of residual interaction. It is shown
that the conception of a zone of reactions can be used for the
determination of the hypersurfaces of a chemical freeze-out and a
sharp kinetic freeze-out.
\end{abstract}

\section{Introduction}
In the collision of nuclei at high energies, a strongly excited
system of interacting particles, a fireball, is formed.
In fact, a fireball is identified with the zone of reactions by such a
definition, i.e. with a space-time region, in which the collisions
of particles occur.
Hence, the zone of reactions must reflect the space-time
characteristics of a fireball, and its study gives information
about the evolution of the interacting system.

While studying the evolution of a fireball,it is important to
know the size of regions, where the majority of various processes
is running.
Depending on a model describing the system, we can distinguish the
regions of the formation of a fireball, its isotropization and
thermalization, the creation of particles, the regions of a chemical
freeze-out and a kinetic one, etc.
This allows one to conditionally select the stages of
evolution of the system and, hence, to obtain the limits of validity
of simple phenomenological models used for the description of the
complicated physical phenomenon, as well as to describe separate
stages of the development of the system in more details.
In particular, the stages of formation ($\tau\sim 0.1$ fm/c) and
thermalization ($\tau\sim 1$ fm/c) are most often described with the
use of microscopic models based on the processes of interaction of
quarks and gluons \cite{Mueller 2000,Arnold 2004,Rebhan 2004}.
To describe the stage of expansion of a dense medium
($1 \le \tau \le 7$ fm/c), the relativistic hydrodynamics is most
often in use, see
\cite{Stoecker-Greiner-1986,Rischke-1998,huovinen-2006,
Russkikh-Ivanov-2006,Romatschke-2009} and references therein.
The further evolution of a hadron gas and the process of kinetic
freeze-out
are covered by kinetic models
\cite{Molnar-Gyulassy-2000,Kisiel-Florkowski-2006,Magas-Csernai-2008}.
As a parameter for the determination of stages of the evolution of
a system, one can take the energy density, mean free path,
frequency of collisions of particles, etc.

In the present work, we use the number of collisions in a given
four-dimensional region of the space-time  as a parameter of the
spatial evolution of the interacting system. Such a quantitative
estimate allows one to define the reaction zone, whose study gives a
possibility to establish the space-time structure of a fireball from
the viewpoint of the interaction intensity at every point of the
space-time. The regions of a fireball can be distinguish by the
interaction intensity which can be characterized by the number of
collisions in a unit volume of the space-time.

Another important question which can be clarified by the study of
the zone of reactions is how the space-time boundary of a fireball
is related to the so-called sharp kinetic freeze-out hypersurface.
Since the kinetic freeze-out is the process of establishment of a
final distribution of particles in the momentum space, the sharp
kinetic freeze-out hypersurface is an imaginary hypersurface,
outside of which there are no collisions between particles of the
system. As a rule, the sharp kinetic freeze-out hypersurface is
defined as such, on which a parameter of the system $L(t,{\bf x})$
takes the critical value $L_c$. That is, the equation of the
hypersurface has form $L(t,{\bf x})=L_c$. As such a parameter, we
may choose the energy density $\epsilon(t,{\bf x})$
\cite{Russkikh-Ivanov-2006,Sollfrank_et.al 1998}, temperature
$T(t,{\bf x})$ \cite{mclerran-1986,Huovinen 2007},
density of particles $n(t,{\bf x})$ \cite{CERES}, etc.
In the present work, we will determine the
sharp kinetic freeze-out hypersurface as a surface bounding the
space-time region, in which almost all collisions between hadrons of
the system happened. That is, we identify the reaction zone boundary
and the sharp kinetic freeze-out hypersurface.

\section{Zone of reactions}
The number of reactions in the given space-time region can be
determined with the use of the distribution function $f(x,p)$. For
example, in the approximation of two-particle collisions, this
function satisfies the Boltzmann equation \cite{groot}
\begin{equation}
p_1^\mu\partial_\mu \, f_1
= \int_2\int_3\int_4 \, W_{12\rightarrow34}\, (f_3\, f_4-f_1\, f_2) \,,
\label{Boltzmann-equation}
\end{equation}
where the right-hand side contains the collision integral.
The quantity $W_{12\rightarrow34}$ is the transition rate which involves
the reaction cross-section and the conservation laws of energy and
momentum, $\int_i\equiv\int {d^3p_i\over (2\pi)^3E_i}$, $f_i\equiv
f(x,p_i)$ are one-particle distribution functions, and
$E_i=\sqrt{m_i^2+{\boldsymbol p}_i^2}$ is the energy of a particle
with momentum ${\boldsymbol p}_i$.

The probability
of a collision of two particles with momenta $\bs p_1$ and $\bs
p_2$ corresponding to the distribution functions $f_1$ and $f_2$,
respectively, is determined at a space-time point $x$ as $\int_3
\int_4 W_{12\rightarrow34}\, f(x,p_1) \, f(x,p_2)$. By integrating
over the momenta of particles, we obtain the frequency of
collisions at the point $x$:
\begin{equation}
\Gamma(x) = \int_1\int_2 \int_3 \int_4 W_{12\rightarrow34}\,
f(x,p_1) \, f(x,p_2)\,.
\end{equation}
Then the number of collisions in the space-time region $\Omega$ is
\begin{equation}
N_{\rm coll}(\Omega) = \int_\Omega d^4x \, \Gamma(x)
=\int_\Omega d^4x \int_1 \int_2  \int_3 \int_4 W_{12\rightarrow34} \,
f(x,p_1)\, f(x,p_2) \, .
\label{coll-number}
\end{equation}
It is seen that the number of collisions in the given space-time
region depends on the frequency of collisions $\Gamma(x)$ which can
be determined in a certain model approximation, e.g., like that in
{\cite{Eskola Ruuskanen 2007,Tomasik Wiedeman 2003,Hung Shuryak
1998}}.
In particular, $\Gamma(x)$ can be determined with the use of
transport models.

\begin{figure}[h!]
    \begin{center}
     \includegraphics[width=17 cm]{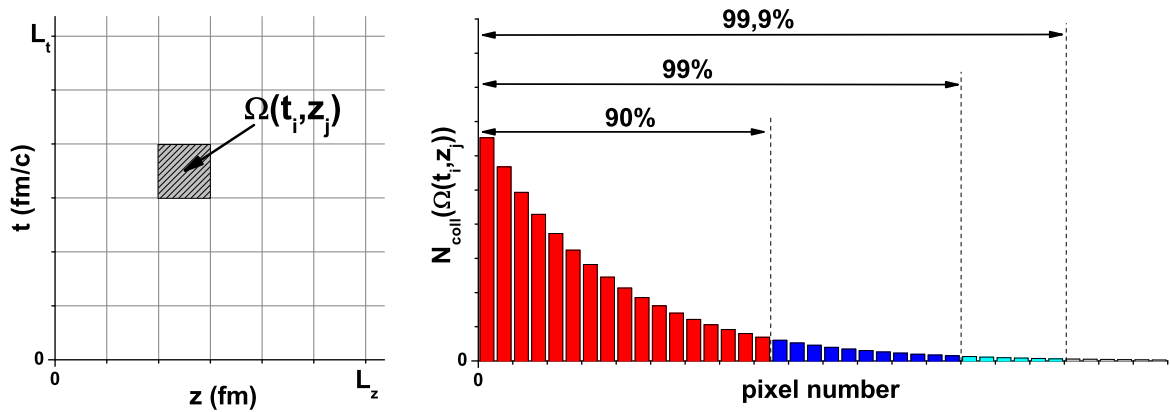}
     \caption{Algorithm of selection of pixels for the determination
     of the zone of reactions.
     }\label{pic.ZoR-method}
    \end{center}
\end{figure}

Let us consider a large space-time region containing a fireball at
its center.
Let this region be so large that the dominant part of all collisions
of hadrons of the system occurs in it.
For example, let this part contain 99.99 \% of all two-particle
reactions and decays of resonances related to this event, i.e., to the
given fireball.
So we obtain a 4-cube of reactions $C_{\rm R}$ with edges
$L_i$, where $i = t,\, x,\, y,\, z$.
In order to determine the zone of reactions, we divide the cube into
separate equal parts (pixels), i.e., elements of the 4-space.
Let $\Omega=\Omega(t,{\bs x})$ be the 4-volume of a pixel with coordinates
of the center of this volume $(t,{\bs x})$.
Totally, we have $N_{\rm pix} = L_tL_xL_yL_z/\Omega$ pixels.
Then, for each four numbers
$(t,{\bs x})$, we can calculate the absolute number of collisions
and decays of resonances in the given pixel $\Omega(t,{\bs x})$
(see Fig.~\ref{pic.ZoR-method}, left panel) by
using, e.g., formula (\ref{coll-number}).
We will determine the absolute number of reactions in each pixel and
place the pixels from left to right by the following hierarchy: from
a pair of pixels, the left pixel is that, in which a larger number of
reactions has occurred.
(In the programming, such a problem is called the ``sorting problem''
and has a standard algorithm of its solution.)
The arrangement of pixels is shown in Fig.~\ref{pic.ZoR-method} (right
panel).
The total area of the whole histogram (area covered by all bins) is equal to
the total number of collisions of all hadrons and decays of resonances
$N_{\rm tot}$ in the 4-cube of reactions $C_{\rm R}$.

It is obvious that the problem of the determination of a space-time
region, where all collisions have occurred, has a statistical character.
Therefore, it is reasonable to search for this region with a certain
precision.
We define ``the zone of reactions'' as a part of the 4-cube of
reactions $C_{\rm R}$, in which a certain share (e.g., 90\% or 99\%)
of all possible processes in the system is running.
A small part of unaccounted reactions, to which we can refer also the
decays of long-lived resonances, happens on such time intervals and
at such distances which exceed the sizes of the 4-cube of reactions
$C_{\rm R}$.

We can calculate a sum of some part of bins depicted in
Fig.~\ref{pic.ZoR-method} (right panel). Namely, we can sum the
areas of bins, by beginning from the left according to the obtained
hierarchy. We recall that the area of each bin gives the number of
reactions in the corresponding pixel. In such a way, we can reach
the value of the sum equal to 90\% of the absolute number of all
reactions $N_{\rm tot}$ (see Fig.~\ref{pic.ZoR-method}).
The hyperspace region which is occupied by the pixels contributing to
this sum gives the zone with the most intense interaction; i.e.,
this region covers 90\% of all hadronic reactions $N_{\rm tot}$.
We call this zone a “hot fireball”. By this algorithm, we will find a
region, where else additionally 9\% of collisions have occurred, and
call it a “cold fireball”.
We also determine a region including else
0.9\% of collisions and call it conditionally a “fireball halo”.
This region corresponds to the residual interaction and is mainly
formed by decays of resonances. We note that the names of the zones
of reactions are conditional and have nothing in common with a real
temperature.

Totally, the three zones include 99.9\% of all hadronic reactions
$N_{\rm tot}$ (see Fig.~\ref{pic.ZoR-method}). As for the available
long-lived resonances, they decay in the large temporal limits and
at distances of the order of hundreds of fm. It is obvious that such
reactions are improbable and are not contained in the basic region
of the evolution of the interacting system. We note that the
above-presented specific numbers (90, 99, and 99.9\%) are quite
conventional and can be replaced by other ones.

\subsection{The results of calculations}

To realize the above-presented algorithm, we use the transport
model UrQMD v2.3 \cite{UrQMD 1998,UrQMD 1999} which allows one to
calculate the density of reactions at every point of the
space-time and to select reactions for the given kind of particles.
In the present calculations, we took a 4-cube of reactions $C_{\rm R}$
with the size of edges $L_i=200$ fm, where $i = t,\, x,\, y,\, z$.

In Figs.~{\ref{pic.ZoR-AGS-ZT-sim}} and {\ref{pic.ZoR-SPS-ZT-sim}},
we show the results of calculations for AGS experiments (Au+Au)
at 10.8A~GeV and for SPS ones (Pb+Pb) at 158A~GeV in the case of
central collisions.
The density of collisions is represented in the coordinates $z$-$t$,
which is the corresponding projection of the reaction zone.
In order to construct this projection of the three-dimensional spatial
pattern onto the $z$ axis, we sum firstly all collisions along the
transverse direction at the fixed coordinates $(t,\,z)$, namely
(see ({\ref{coll-number})),
\begin{equation}
\widetilde N_{\rm coll}(\widetilde \Omega(t,z))
=  \int dx\, dy\, N_{\rm coll}(\Omega(t,x,y,z))\, .
\label{coll-number-tz}
\end{equation}
That is, we put a number of collisions $\widetilde N_{\rm coll}$ in
correspondence to the pixel $\widetilde \Omega(t,z)$ with the
coordinates $(t,\, z)$.
Then we construct the hierarchy of pixels $\widetilde \Omega(t,z)$
according to the sorting
algorithm, where the basic quantity is now the number of collisions
$\widetilde N_{\rm coll}$ (see Fig.~\ref{pic.ZoR-method}).
%
\begin{figure}[h!]
   \begin{center}
     \includegraphics[width=15cm]
     {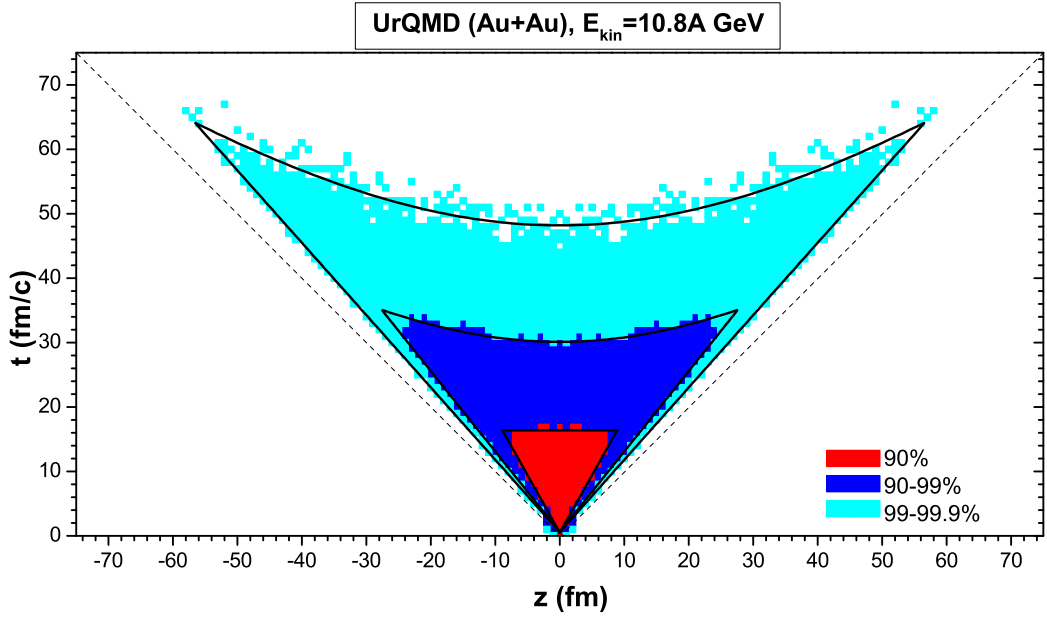}
     \caption{Projection of the reaction zone on the $z-t$ plane
     [AGS (Au+Au at 10.8A~GeV) conditions].}
\label{pic.ZoR-AGS-ZT-sim}
   \end{center}
\end{figure}
\begin{figure}[h!]
    \begin{center}
     \includegraphics[width=15cm]
     {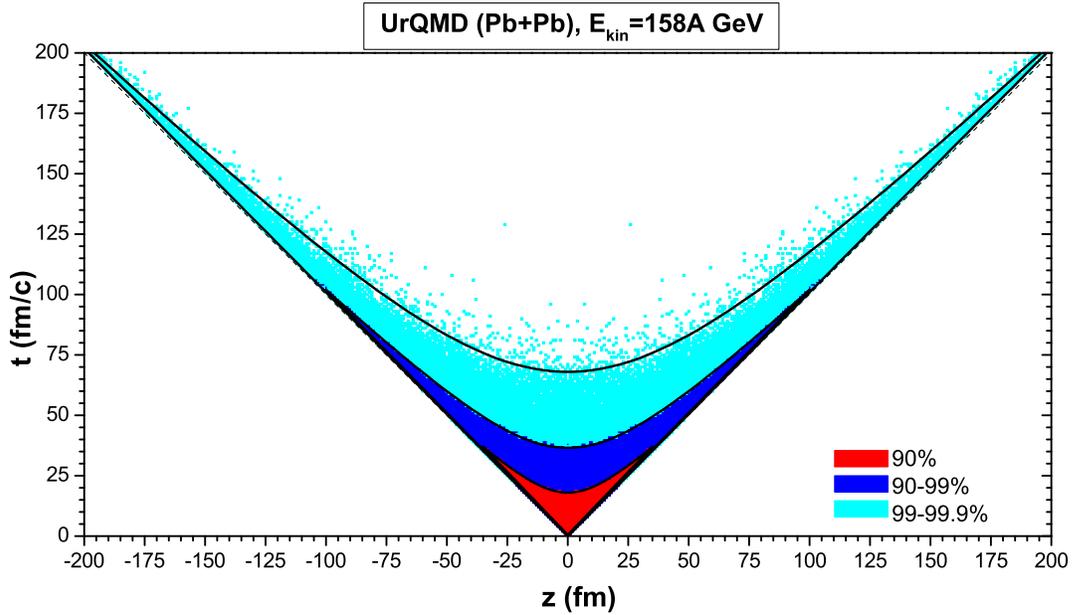}
     \caption{Projection of the reaction zone on the $z-t$ plane
     [SPS (Pb+Pb at 158A~GeV) conditions].   }
\label{pic.ZoR-SPS-ZT-sim}
    \end{center}
\end{figure}
Further, like the general case where we dealt with the 4-cube of
reactions $C_{\rm R}$, we sum, in succession, the areas of
bins from left to right, i.e., we determine the total number of reactions.
When we obtain the value of the sum equal to 90\% of the absolute
number of all reactions $N_{\rm tot}$, we arrange the pixels
contributing to this sum on the plane in the correspondence to their
coordinates $(t,\,z)$.
We mark the zone occupied by these pixels with red color and call it
a “hot fireball” in the $z$-$t$ projection (see
Figs.~{\ref{pic.ZoR-AGS-ZT-sim}} and {\ref{pic.ZoR-SPS-ZT-sim}}).
Analogously, we gather additionally a sum of the areas of bins which
gives 9\% of the total number of all reactions $N_{\rm tot}$
and name this zone a “cold fireball” in the $z$-$t$ projection.
The zone of a “cold fireball” in Figs.~{\ref{pic.ZoR-AGS-ZT-sim}} and
{\ref{pic.ZoR-SPS-ZT-sim}} is drawn with blue color.
By continuing to move along the hierarchical structure of pixels from
left to right and by gathering a sum of the areas of their bins, i.e.
the numbers of collisions corresponding to pixels, we determine those
pixels which give, in total, else 0.9\% of the total number of all
reactions $N_{\rm tot}$.
By arranging them on the same plane in the correspondence to the
coordinates $(t,\,z)$ of each pixel, we obtain a zone, name it a
“fireball halo” in the $z$-$t$ projection, and mark it with cyan-blue
color.
That is, the three zones cover 99.9\% of the total number
$N_{\rm tot}$ of all hadronic reactions.

Two competing processes
coexist in the system arising in a collision of relativistic ions.
On the one hand, as a result of the collision of nuclei, their initial
kinetic energy of a longitudinal motion is transformed into the
energy of created secondary particles, e.g. $\pi$-mesons, and into the
energy of transverse motion.
The density of particles increases; as a result, the probability of
collisions of particles grows.
On the other hand, it is the rapidly expanding system, which is
accompanied by a drop of the density of particles.
Thus, we determine the region of intense interaction,
in which the process of creation of new particles compensates a
decrease in the density due to the spreading of the system.

The analysis performed with the use of UrQMD indicates that the zone
of a hot fireball contains all inelastic reactions running in the
system, 98\% and 99\% of all inelastic reactions for AGS and SPS,
respectively.
Moreover, the numbers of decays and fusions of particles in this zone
are practically the same.
If we accept the viewpoint that the chemical freeze-out happens with
the termination of inelastic reactions \cite{Heinz 2001}, then,
according to this definition, the chemical freeze-out hypersurface can
be associated with the boundary of the zone of a hot fireball.

The zone of a cold fireball includes 9$\%$ of all reactions
in the system.
Their greater part consists of decays of resonances, 49\% and 56\% of
all reactions in this zone for AGS and SPS, respectively.
In the residual interaction zone (a fireball halo) marked with
cyan-blue color, else 0.9\% of reactions run.
The decays of resonances in the residual interaction zone contribute
more than those in the zone of a cold fireball, 51\% and 64\%
of all reactions in this zone for AGS and SPS, respectively.
Together the zone of a cold fireball and the fireball halo form a
space-time region occupied by a hadron-resonance gas.
In this region, the reactions are represented by the processes of
decays of resonances, fusion of particles, and elastic scattering.
In this case, the number of the decays of resonances exceeds the total
number of other processes.
Since the processes of fusion run earlier than the processes of decay,
which dominate in these two regions, we may assert that the greater
part of particles in these regions leaves the system as a result of
the decays of resonances.

In Figs. {\ref{pic.ZoR-AGS-ZT-sim}} and {\ref{pic.ZoR-SPS-ZT-sim}},
we present the projections of the reaction zones on the $z$-$t$ plane.
It is seen that the interacting system has a comparatively high
lifetime -- a hot fireball (in the on-line representation, the region
is marked by red color) decays completely only in the time
intervals of the order of 16 fm/c for AGS and 50 fm/c for SPS.
A cold fireball (in the on-line representation, the region is marked
by blue color) lives for $32\div 34$~fm/c for AGS and
$90\div 100$~fm/c for SPS, respectively.
Black lines show the approximation of the boundaries of the reaction
zones.

By definition, the zones of hot and cold fireballs together contain
99\% of all reactions.
Moreover, the greater part of reactions outside of these zones
consists of decays of resonances.
Thus, the hypersurface which bounds the region of a cold fireball
(in the on-line representation, it is the boundary between the blue
and cyan-blue regions) can be taken as the sharp freeze-out
hypersurface.
For both collision energies of nuclei under consideration, the upper
part of the boundary of the reaction zone,
i.e. $t=t_{\rm B}(z)$, is a space-like hyperbola and takes the
form of a constant proper-time surface to within some factor, namely
$t_{\rm B}(z)=A\sqrt{\tau^2+z^2}$, where $A$=0.75, $\tau$=64~fm/c for
AGS energies and $A$=0.95, $\tau$=38~fm/c for SPS energies,
respectively (see Figs. {\ref{pic.ZoR-AGS-ZT-sim}} and
{\ref{pic.ZoR-SPS-ZT-sim}}).
The lower time-like surface bounding a cold fireball has the form of a
straight line $t_{\rm B}(z)=t_0+{1\over v}z$, where $t_0$ is close to
zero, and $v$=0.8 for AGS energies and $v$=0.98 for for SPS energies.
We note that, for SPS energies $E_{\rm kin} = 158$A~GeV, the time-like
surfaces bounding all three zones of a fireball practically coincide
with one another and are close to the light cone.
This does not else occur for AGS energies
$E_{\rm kin} = 10.8$A~GeV, where the time-like
boundaries of the reaction zones differ significantly from one
another and the light cone (see Figs. \ref{pic.ZoR-AGS-ZT-sim},
\ref{pic.ZoR-SPS-ZT-sim}).

It is useful to represent the reaction zones in the coordinates
$\eta$ and $\tau$,
\begin{equation}
\eta={1\over 2}\ln{\left({{t+z}\over{t-z}} \right) } \,,
\qquad
\tau=\sqrt{t^2-z^2}
\, ,
\label{eta-tau}
\end{equation}
where $\eta$ is the longitudinal coordinate rapidity, and $\tau$ is
the proper time. In Figs. {\ref{pic.ZoR-AGS-etatau-sim}} and
{\ref{pic.ZoR-SPS-etatau-sim}}, we present the reaction zones in
these coordinates, respectively, for the collision energies of
nuclei on AGS and SPS.

In order to construct projections of the reaction zone in the
coordinates  $\eta$-$\tau$, we determine the number of reactions
$\widetilde N_{\rm coll}\big(\widetilde \Omega (\tau,\eta)\big)$ in
each pixel $\widetilde \Omega (\tau,\eta)$.
Then, by using the above-presented
algorithm, we construct the reaction zones. In correspondence with
the previous figures {\ref{pic.ZoR-AGS-ZT-sim}} and
{\ref{pic.ZoR-SPS-ZT-sim}}, we mark the zone of a hot fireball,
where 90\% of all reactions run, by red color. The zone of a cold
fireball and the fireball halo are indicated by blue and cyan-blue
colors, respectively. In Figs. {\ref{pic.ZoR-AGS-ZT-sim}} and
{\ref{pic.ZoR-SPS-ZT-sim}}, black lines show the upper boundaries of
the reaction zones in their representation in the coordinates
$\eta$-$\tau$. That is, these lines are a result of the
transformation $(t,\,z) \to (\tau,\, \eta)$ of the space-like
hyperbolas $t=A\sqrt{\tau^2+z^2}$ which correspond to boundaries of
all three stages of the evolution of the fireball.
%
\begin{figure}[h!]
    \begin{center}
     \includegraphics[width=15cm]
{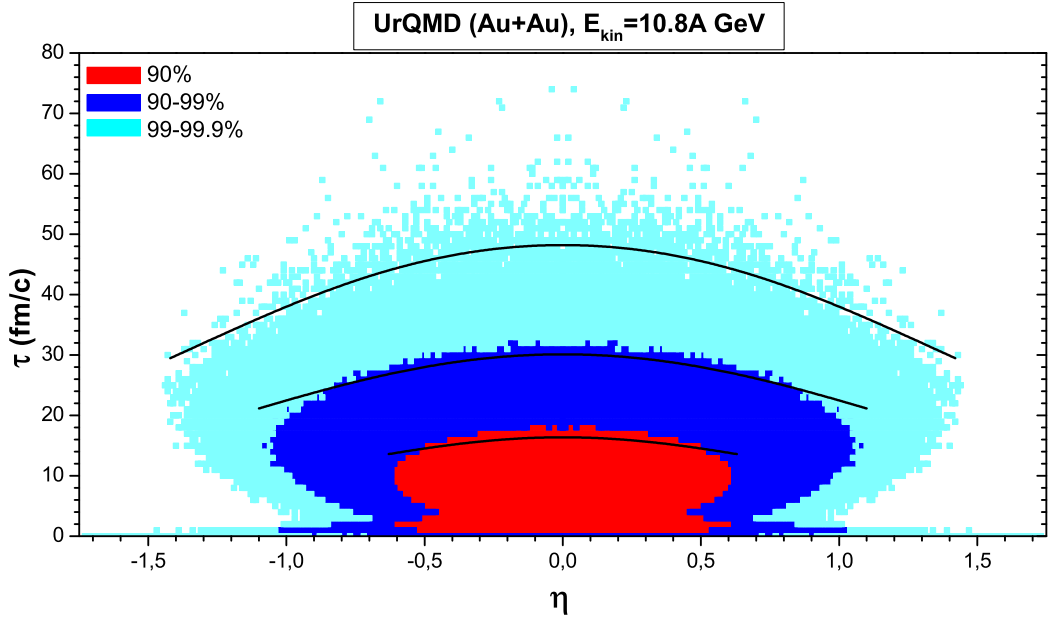}
\caption{Projection of the reaction zone in the $\eta$-$\tau$
coordinates.  }
\label{pic.ZoR-AGS-etatau-sim}
    \end{center}
\end{figure}
\begin{figure}[h!]
    \begin{center}
     \includegraphics[width=15cm]
{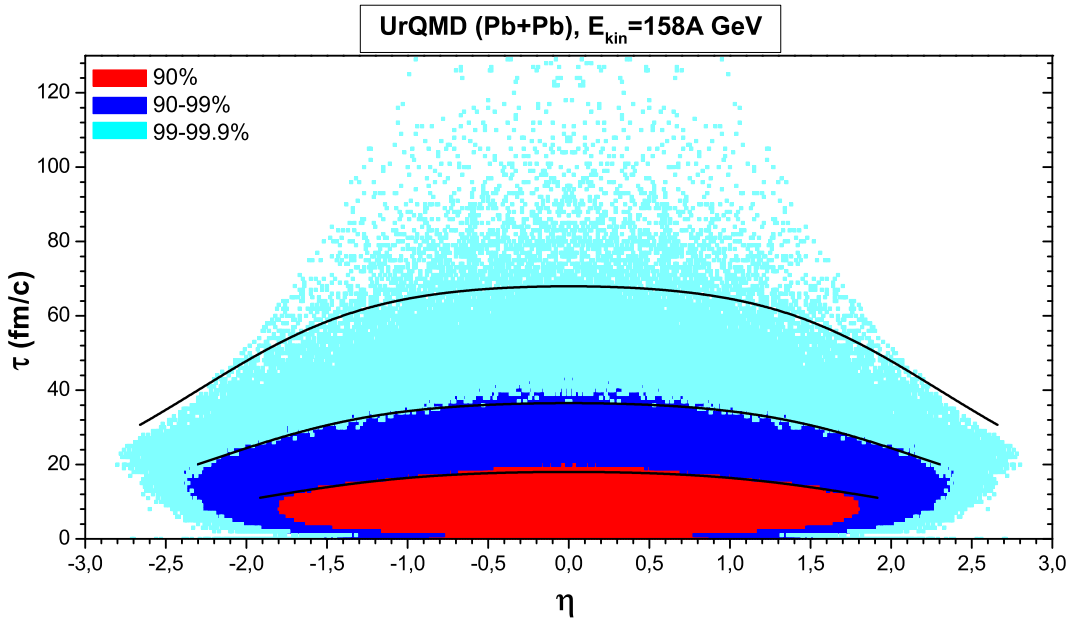}
\caption{Projection of the reaction zone in the $\eta$-$\tau$
coordinates. }
\label{pic.ZoR-SPS-etatau-sim}
    \end{center}
\end{figure}

The presentation of the reaction zone in the coordinates $\eta$ -
$\tau$ allows one to see the rapidity region, where the processes
of collisions of particles run.
In particular, Figs.
{\ref{pic.ZoR-AGS-etatau-sim}} and {\ref{pic.ZoR-SPS-etatau-sim}}
show that the maximum intervals of the coordinate longitudinal
rapidity $\Delta \eta$ coincide with the intervals of the momentum
rapidity $\Delta y = 3.2$ for the energy
$E_{\rm kin} = 10.8$A~GeV on AGS and $\Delta y = 5.8$ for the energy
$E_{\rm kin} = 158$A~GeV on SPS, respectively. For
both energies, the reaction zones differ significantly along the
$\eta$ axis, but their sizes differ slightly along the $\tau$
axis. Indeed, the size of the zone of a hot fireball near $\eta=0$
for reactions with participation of all hadrons on SPS is by 10\%
larger than that of the corresponding region on AGS
($\Delta\tau_{\rm hot} = 17$~fm/c for AGS and $\Delta\tau_{\rm
hot} = 19$~fm/c for SPS, respectively).
The spreading of the reaction zone along the coordinate rapidity axis
at the transition from the AGS- to SPS-energies is a result of the
increase in the initial energy of particles.
As was shown in
{\cite{Anch.Yezhov.Musk. 2009p1,Anch.Yezhov.Musk. 2009p2}}, the
edges of the momentum rapidity distribution $dN/dy$ of nucleons are
formed by those nucleons which participated only in several
collisions.
We may assume that the initial direction of motion
and the initial velocity are almost invariable after the first
collision at the eikonal scattering, and, therefore, the momentum
rapidity of a scattered particle is close to the initial one.
Just such particles scatter then elastically on one another and form
the reaction zone at great values of $\eta$ (what corresponds to big
values of $t$ and $z$).

At the same time, the size of a reaction zone along the $\tau$ axis
at $\eta=0$ remains approximately invariable, as the kinetic energy
increases by more than 15 times. This can be related to
peculiarities of the behavior of the system in the central region of
the momentum rapidity, which corresponds, undoubtedly, to the $\eta=0$
region of the coordinate rapidity.
For example, it was shown in
{\cite{Anch.Yezhov.Musk. 2009p1,Anch.Yezhov.Musk. 2009p2}} that the
spectrum of nucleons in the region near $y=0$ is determined by the
thermal distribution or a distribution close to it irrespective of
the initial collision energy. Therefore, the course of reactions in
this region for identical nuclei and the same centrality will be
invariable, i.e., it will slightly depend on the energy.

\section{Conclusions}
Thus, the proposed algorithm of the determination of the reaction zone
allows one to get information about the space-time structure of an
interacting system which is created in the collision of
relativistic ions.

By taking the number of collisions which have occurred in unit
volume of a space-time region as the interaction intensity degree,
we can separate the following parts of a fireball which characterize
its  evolution (see Figs. 2--5): 1) Region of a ``hot'' fireball,
where 90\% of all hadronic reactions, i.e. 0.9$N_{\rm tot}$, have
occurred; as has been shown, this region of intense interaction
contains all inelastic collisions (in the on-line presentation, this
zone is marked by red color); 2) Region of a ``cold'' fireball,
where 9\% of all hadronic reactions, i.e. 0.09$N_{\rm tot}$, have
occurred (in the on-line presentation, this zone is marked by blue
color); 3) Region of a fireball halo, where 0.9\% of all hadronic
reactions, i.e. 0.009$N_{\rm tot}$, have occurred (in the on-line
presentation, this zone is marked by cyan-blue color). Two last
zones together are a space-time region containing the
hadron-resonance gas, and the reactions in this region are mainly
presented by decays of resonances.

The hypersurface separating the zones of a hot fireball and a cold
one  can be associated with the chemical freeze-out hypersurface (we
assume that the chemical freeze-out occurs, when the inelastic
reactions are completed \cite{Heinz 2001}).

Let us define that the kinetic freeze-out hypersurface separates
the region, where the collisions between particles of the fireball
are running, from the region, where there are no collisions. Then
the hypersurface including the first two zones, i.e. the boundary
between the zone of a cold fireball and the fireball halo, can be
interpreted as a sharp kinetic freeze-out hypersurface. In the
coordinates $(t,\,z)$, the space-time part of this hypersurface is a
hyperbola and has form $t=A\sqrt{\tau^2+z^2}$, where $A$=0.75,
$\tau$=64~fm/c for the AGS-energy $E_{\rm kin} = 10.8$A~GeV and
$A$=0.95, $\tau$=38~fm/c for the SPS-energy $E_{\rm kin} = 158$A~GeV.

\section*{Acknowledgements}
The authors thank L.~McLerran and V.~Magas for the discussion.
The authors grateful P.~Romatschke for the fruitful discussions
of the present work and the valuable comments.
D.~Anchishkin was partially supported by the program
\textquotedblleft Fundamental properties of physical systems under
extreme conditions\textquotedblright\ (Section of the physics and
astronomy of the NAS of Ukraine).


\end{document}